\documentclass[aps,prl,twocolumn,amsmath,amssymb]{revtex4-1}
\usepackage{graphicx}
\usepackage{tabularx}

\usepackage{color}

\newcommand{\Ang}{\,\mathrm{\AA}}

\def\eV{\,\textrm{eV}}

\def\cis{{\it cis}}
\def\trans{{\it trans}}

\begin{document}

\title{
Multi-Control Over Graphene-Molecule Hetereo-Junctions
}
\author{Yun-Peng Wang,$^{1,2} $ J. N. Fry,$^1$  and Hai-Ping Cheng$^{1,2}$ } 
\email[Corr. author: Hai-Ping Cheng,  ]{ hping@ufl.edu }
\affiliation{$^1$Department of Physics, University of Florida,  Gainesville, Florida 32611, USA \\
             $^2$Quantum Theory Project, University of Florida, Gainesville, Florida 32611, USA}

\begin{abstract}

The vertical configuration is a powerful tool recently developed experimentally 
to investigate field effects in quasi 2D systems.
Prototype graphene-based  vertical tunneling transistors 
can achieve an extraordinary control over current density utilizing gate voltages.
In this work we study theoretically vertical tunneling junctions 
that consist of a monolayer of photo-switchable aryl-azobenzene molecules 
of sandwiched between two sheets of graphene.
Azobenzene molecules  transform between \trans{} and \cis{} conformations upon 
photoexcitation, thus adding a second knob 
that enhances control over physical properties of the junction.
Using first-principles methods within the density functional framework, 
we perform simulations with the inclusion of 
field effects for both \trans{} and \cis{} configurations.
We find that the interference of interface states 
resulting from molecule-graphene interactions at the Fermi energy 
introduces a dual-peak pattern in the transmission functions 
and dominates the transport properties of gate junctions, 
shedding new light on interfacial processes. 

\end{abstract}
\maketitle

{\it Introduction.}
In miniaturizing metal-oxide-semi\-conductor field-effect transistors (MOSFETs) to sub-$ 5 \, \textrm{nm} $ size, one encounters severe challenges in preserving system performance and reliability \cite{Nature.530.144}, and one must rely on new physics emerging from nano-scale material systems to overcome this obstacle. The vertical integration of tunneling field effect transistors (TFETs) based on the stacking of two-dimensional (2D) layered materials has been explored as an alternative transistor architecture in recent years \cite{Adv.Mater.28.2547,Nature.526.91,Science.335.947,NNANO.8.100,NanoLett.15.428}. Vertical TFETs stacking layers of graphene with 2D insulators and semiconductors such as hexagonal BN and transition-metal dichalcogenides have been demonstrated to exhibit extraordinary control of the tunneling current via gate voltages \cite{Science.335.947,NNANO.8.100,PhysRevB.91.245307}. Equally if not more important, the vertical configuration introduces a technique to approach fundamental physics processes at interfaces, an emergent area in materials, chemical, and condensed matter physics \cite{Sci.Rep.4.6220,Nat.Comm.6.6181,ACSNano.10.3816}.

When the thickness of a system is of nano-scale, interfacial physics governs the electronic, magnetic and transport properties. For instance the $h$-BN\,$|$\,graphene interface is electrostatically sharp \cite{PhysRevB.91.245307}, while the dielectric response of $ \textrm{WS}_2 $ is greatly affected by interface with graphene \cite{XGLi-WS2}. Interface physics is in general not limited to nano-scale materials; interfaces between two bulk systems or between a bulk material and vacuum or air can also trigger emergent phenomena. Examples include the two-dimensional electron gas at a $ \textrm{LaAlO}_3 \, | \, \textrm{SrTiO}_3$ interface \cite{Nature.427.423} and the protected surface states of the celebrated topological insulators \cite{RevModPhys.83.1057}. Nano-structured 2D systems, which have bulk 2D extent but are finite in the third dimension, provide unique opportunities for investigating and understanding interfacial processes.

The subject of this study is TFETs based on graphene and organic materials. Recent advances in fabricating hybrid nano-scale heterostructures have made it possible to incorporate organic molecules in the design of vertical TFETs. Covalent bonds between molecules and graphene can form via dediazoniation \cite{JACS.117.11254,JACS.131.1336}. It was proposed that such bonds enhance the flexibility of graphene\,$|$\,molecule\,$|$\,graphene junctions \cite{Nat.Comm.4.1920}. The gate-voltage tunable transport through graphene\,$|$\,molecule interfaces has been attributed to modulation of the Schottky barriers at the interface \cite{NatComm.5.5162,Adv.Funct.Mater.25.2972,ACSNano.9.5922}.
Organic molecules offer the advantage over inorganic materials that their properties can be modified by substituting various ligands; in particular, photo-switchability can be incorporated by inserting azobenzene ligands \cite{Langmuir.14.4955,Langmuir.17.4593,NanoLett.4.551,Nat.Comm.4.1920,PhysRevB.89.115415} that can switch between {\trans} or {\cis} conformations. Electronic, magnetic and photoelectronic properties of molecules on graphene have been studied previously \cite{PhysRevB.89.245447,PhysRevB.90.125447,JPCC.119.22357}. In this work we investigate theoretically the interface states at graphene\,$|$\,azobenzene(containing)\,$|$\,graphene nano-junctions and the emerging phenomena when gate voltage is turned on. Combined with optical control over the \trans{} and the \cis{} forms of azobenzene molecules, we propose a multi-control charge transport process and the underlying physical origin.

\begin{figure}[b]
\begin{center}
\includegraphics[width=\linewidth]{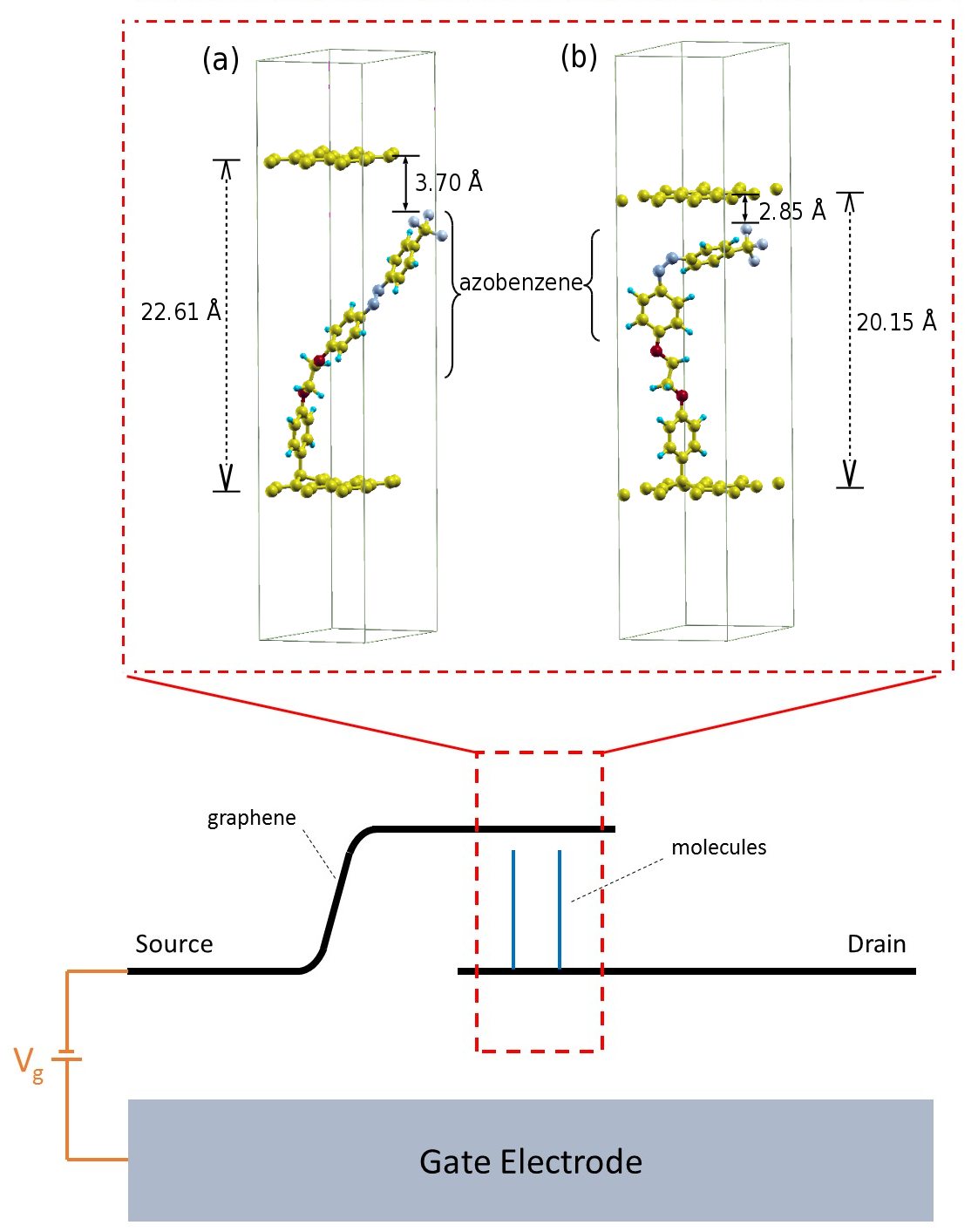}
\caption{
\label{fig:transport-structure}
(color online) 
Schematics of graphene$\,|\,$molecule$\,|\,$\-graphene tunneling field effect transistors (TFETs).
A gate voltage $V_g$ is applied to the gate electrode below.
The molecule can be in (a) \trans{} and (b) \cis{} conformations.
}
\end{center}
\end{figure}

{\it Computational Methods.}
Calculations were carried out using density functional theory (DFT). The geometries of graphene\,$|$\,\trans\,$|$\,graphene (Gr$\,|\,${\trans}$\,|\,$Gr) and graphene\,$|$\,\cis\,$|$\,graphene (Gr$\,|\,${\cis}$\,|\,$Gr) were fully optimized using the VASP package \cite{CMS.6.15,PhysRevB.54.11169}. The residual forces on atoms are smaller than $ 0.01 \eV/\Ang$ after relaxation. The weak interactions between molecules and the top graphene layer were treated by using the van der Waals density functional \cite{JPCM.22.022201,PhysRevB.83.195131}. Within the generalized gradient approximation parameterized by Perdew-Burke-Ernzerhof \cite{PhysRevLett.77.3865}, the electronic structure and transport properties were calculated using the SIESTA package, \cite{SIESTA} based on Troullier-Martins norm-conserving pseudopotentials \cite{PhysRevB.43.1993} and a pseudo-atomic orbital basis set.
The electrostatic interaction was treated by the effective screening medium (ESM) approach \cite{PhysRevB.73.115407} that solves the Hartree potential with non-periodic boundary conditions, and we recently adopted the ESM method to simulate gate effects \cite{PhysRevB.91.245307,PhysRevB.94.165428}. Positive gate voltages $V_g$ induce extra electrons into the Gr$\,|\,$molecule$\,|\,$Gr system. A gate electric field of $ \epsilon_0 E_g = 0.01\, \textrm{C/m}^2 $ 
increases the areal electron density by  $ 6.24 \times 10^{12}\, {\rm cm}^{-2}$.

The molecule-graphene heterogeneous TFET is schematically shown in Fig.~\ref{fig:transport-structure}. Azobenzene containing molecules 4-(10-(4-((4-(trifluoromethyl)-phenyl)diazenyl)phenoxy)ethenyloxy)benzene are anchored at the bottom graphene layer by a C-C bond and the contact with the top graphene layer is of van der Waals nature. The leftmost and rightmost graphene regions serve as source and drain. 

The curved part of the left graphene sheet follows the description in Ref.~\onlinecite{PhysRevLett.108.096601}, where it was shown that electrons encounter little scattering. This configuration mimics the experimental configuration, and gate voltages are guaranteed to induce the same potential shift on source and drain. The equilibrium Green's function and transmission probability were calculated from the Hamiltonian using well-established procedure \cite{PhysRevB.65.165401}.

\begin{figure}[t]
\begin{center}
\includegraphics[width=\linewidth]{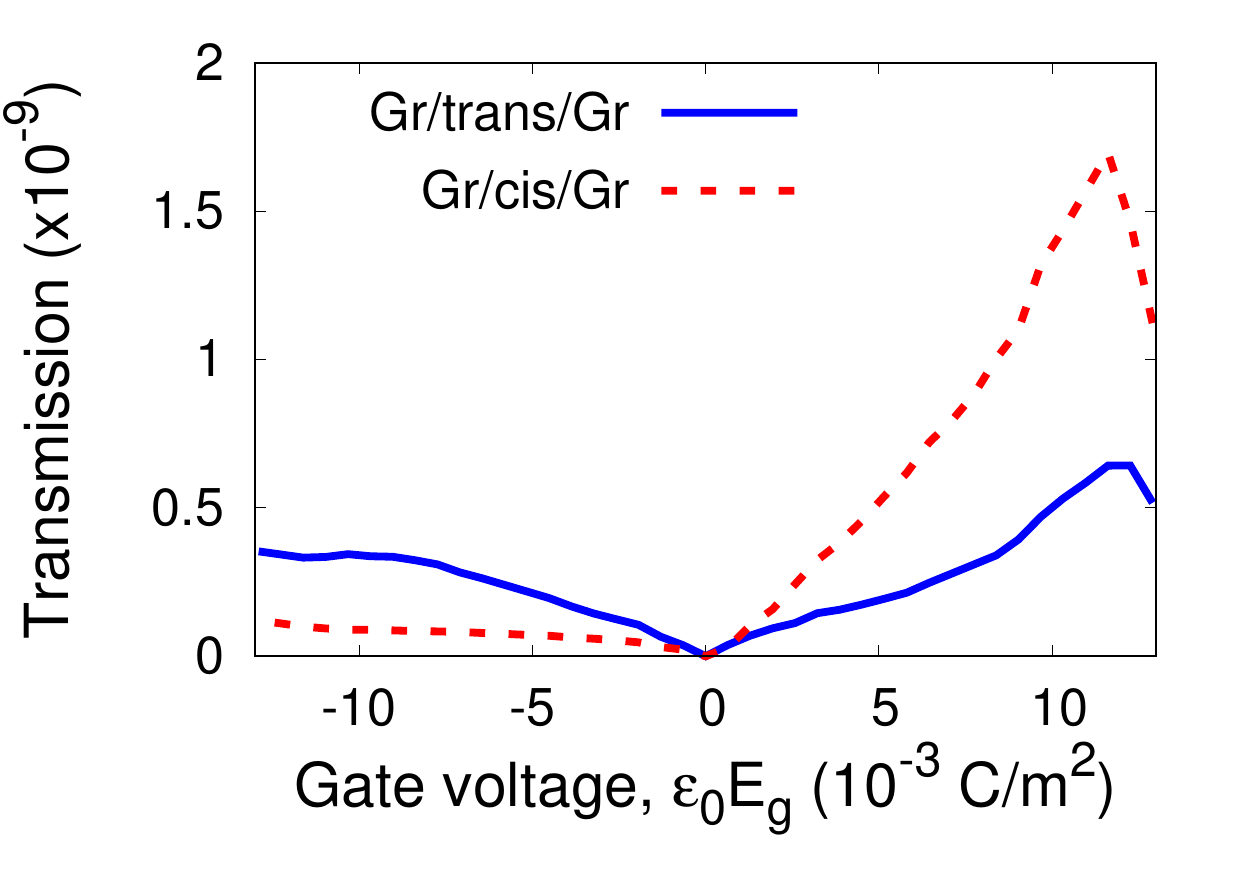}
\caption{
\label{fig:transmission_at_Ef}
(color online) 
Transmission at the Fermi energy as a function of gate voltage 
for Gr$\,|\,${\trans}$\,|\,$Gr and Gr$\,|\,${\cis}$\,|\,$Gr junctions.
}
\end{center}
\end{figure}


\begin{figure}[b]
\begin{center}
\includegraphics[width=\linewidth]{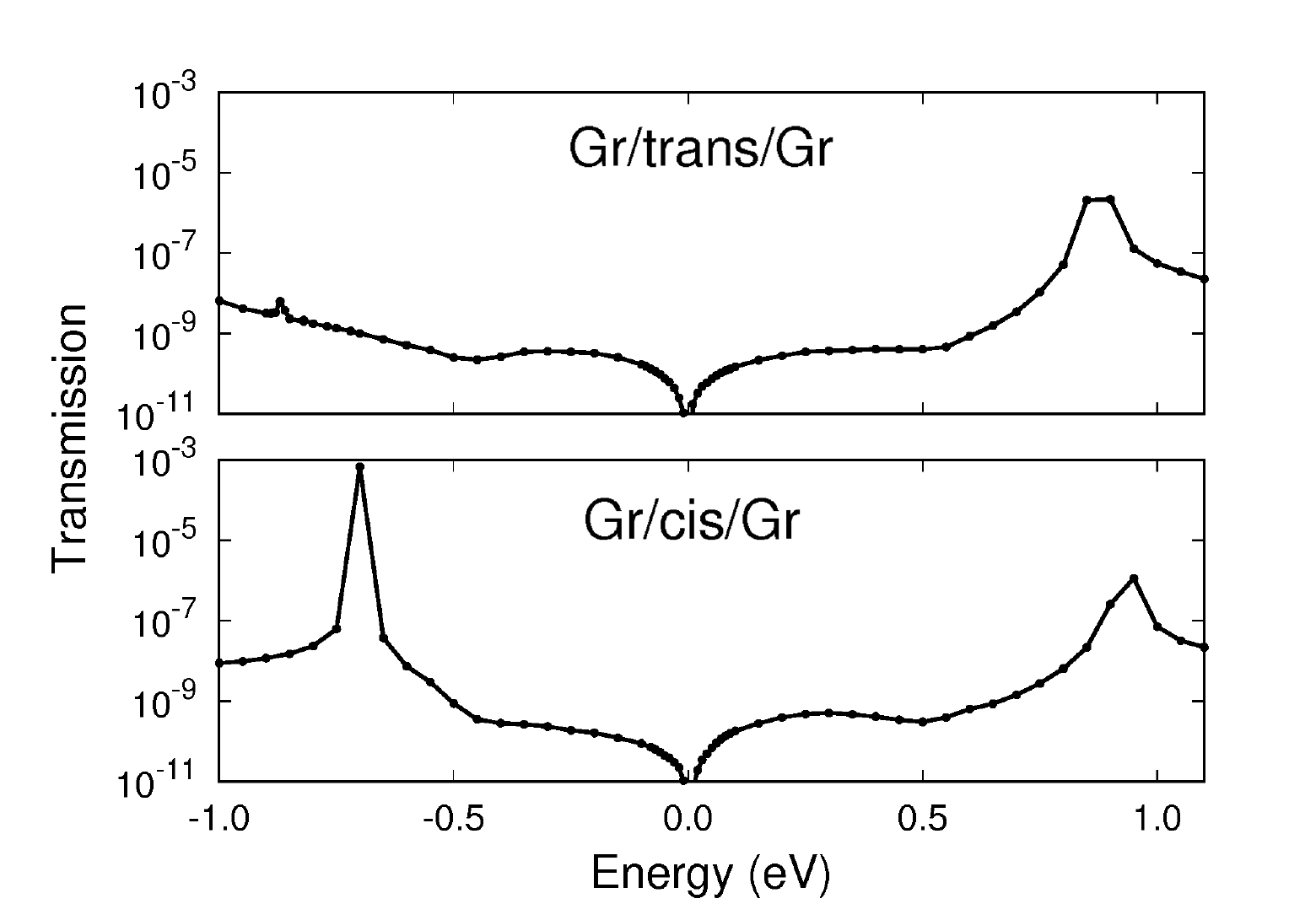}
\caption{
\label{fig:transmission_0}
(color online) 
Transmission as a function of energy for (top) Gr$\,|\,${\trans}$\,|\,$Gr and
(bottom) Gr$\,|\,${\cis}$\,|\,$Gr junctions under zero gate voltage.
}
\end{center}
\end{figure}

\begin{figure} 
\begin{center}
\includegraphics[width=0.8\linewidth]{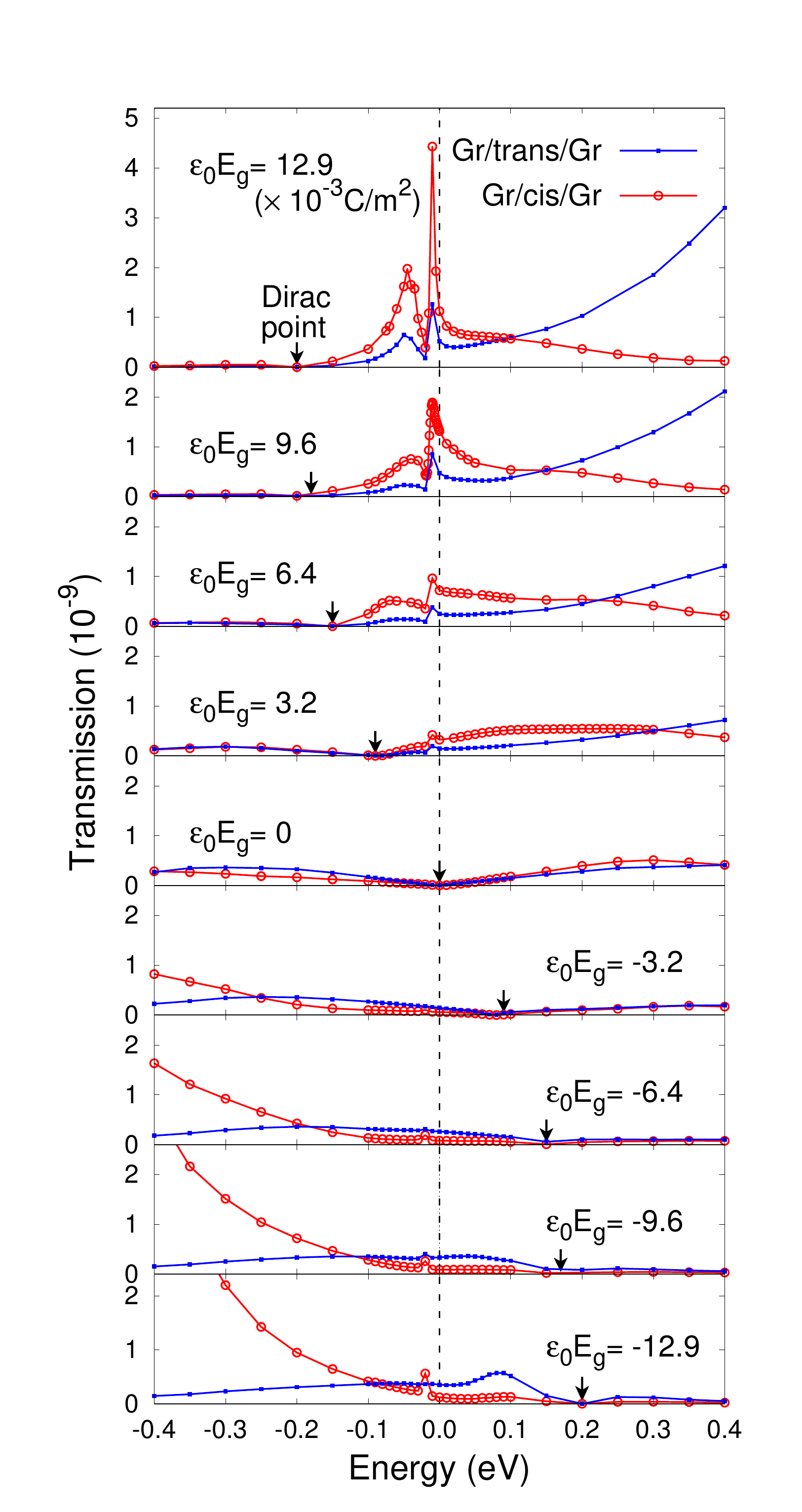}
\caption{
\label{fig:transmission_vg}
(color online) 
Transmission curves under different gate voltages of Gr$\,|\,${\trans}$\,|\,$Gr 
and Gr/\cis/Gr junctions.
The position of the Dirac point of the graphene lead is denoted by arrows.
}
\end{center}
\end{figure}

{\it Results.}
Transmission of Gr$\,|\,$molecule$\,|\,$Gr junctions depends on the conformation ({\trans} and {\cis}) of the molecule, and can be modulated by gate voltage, as shown in Fig.~\ref{fig:transmission_at_Ef}. The Gr$\,|\,${\trans}$\,|\,$Gr junction has a transmission almost symmetric with respect to negative and positive gate-voltage, while transmission of the Gr$\,|\,${\cis}$\,|\,$Gr junction is much enhanced by positive gate voltages. With no gate voltage, the different conductance of azobenzene containing molecules with {\trans} and {\cis} conformations was attributed to the difference in their lengths \cite{Nat.Comm.4.1920}, different couplings to electrodes \cite{PhysRevLett.92.158301}, and different tunneling pathways \cite{PhysRevB.86.035444}. The astonishing gate effect as seen in Fig.~\ref{fig:transmission_at_Ef} on the transmission reveals new physics that requires explanation.

Electronic transport through molecules is usually governed by molecular orbitals, especially the highest occupied molecular orbital (HOMO) and the lowest unoccupied molecular orbital (LUMO), those closest to the Fermi energy. Transmission by a molecular orbital exhibits a Lorentzian peak centered at approximately its energy and with width proportional to the strength of its coupling to electrodes \cite{PhysRevB.35.9805,PhysRevB.38.3913}. Both Gr$\,|\,${\trans}$\,|\,$Gr and Gr$\,|\,${\cis}$\,|\,$Gr junctions  show two peaks in their transmission curves at zero gate voltage (Fig.~\ref{fig:transmission_0}). Peaks at around $ -0.7 \eV $ correspond to the molecular HOMO, and those near $ 1.0 \eV $ are attributed to the LUMO. The peak corresponding to the HOMO in \trans{} junctions is narrow, which indicates a weaker coupling between this orbital and graphene electrodes. The dips in the transmission curves near the Fermi energy are the results of the small density of states in graphene electrodes. We also examined the evolution of the transmission curves of Gr$\,|\,${\trans}$\,|\,$Gr and Gr$\,|\,${\cis}$\,|\,$Gr junction under different gate voltages (Fig.~\ref{fig:transmission_vg}) as a function of energy. Note that the energy range in here is smaller than in Fig.~\ref{fig:transmission_0}, and transmission peaks due to molecular LUMO and HOMO do not appear in this energy window.

The Dirac point of graphene electrodes shifts along with the gate voltage (arrows in Fig.~\ref{fig:transmission_vg}). Peaks in transmission close to the Fermi energy are responsible for the Gr$\,|\,${\trans}$\,|\,$Gr junction showing a higher transmission than the Gr$\,|\,${\cis}$\,|\,$Gr junction at positive gate voltage (Fig.~\ref{fig:transmission_at_Ef}). Near the Fermi energy, these emergent peaks cannot be explained by molecular orbitals, which are far away from the Fermi energy. 



\begin{figure} 
\begin{center}
\includegraphics[width=\linewidth]{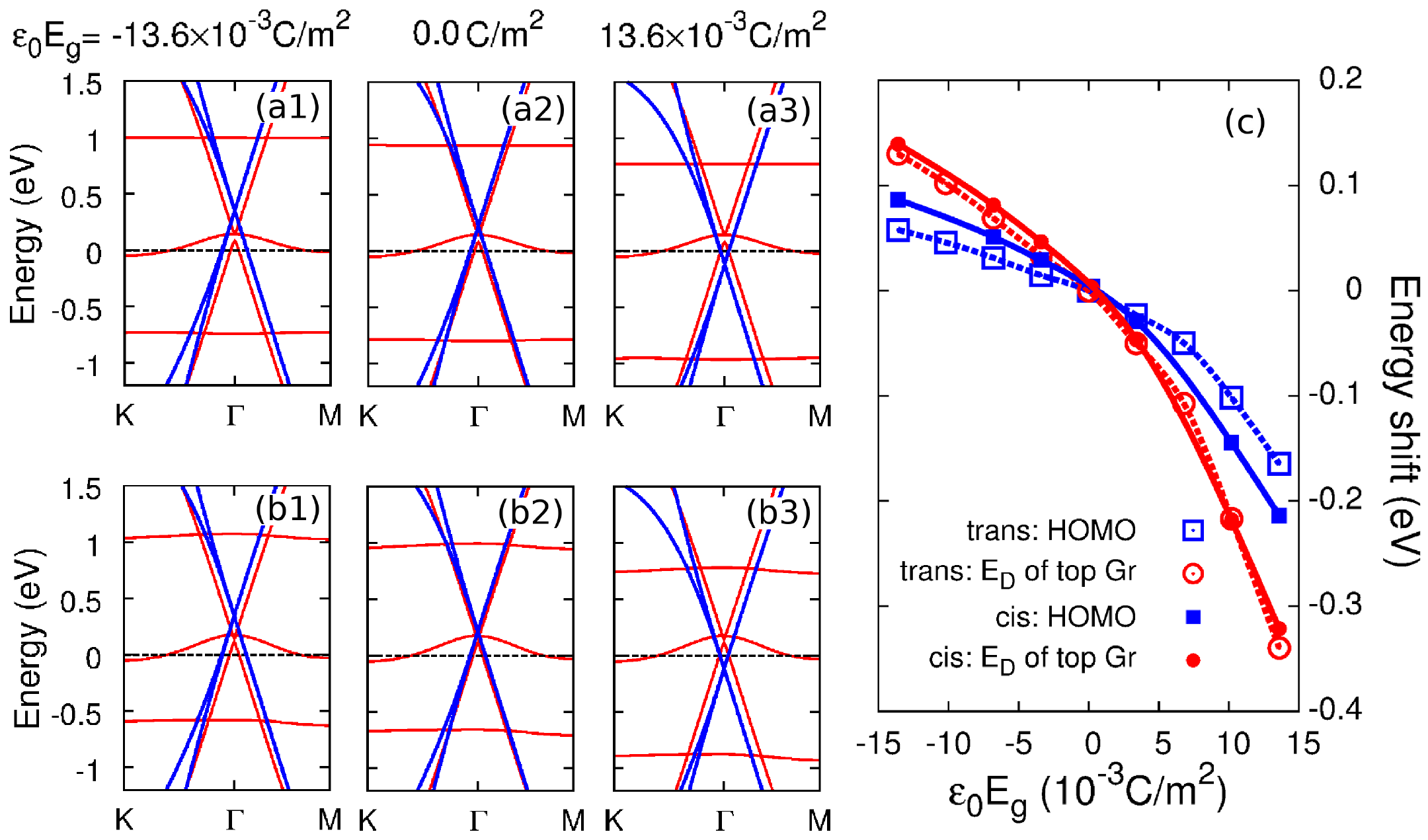}
\caption{
\label{fig:bands}
(color online) 
Band structure of (a1)--(a3) Gr$\,|\,${\trans}$\,|\,$Gr and (b1)--(b3) Gr$\,|\,${\cis}$\,|\,$Gr junctions for gate voltages of
(a1, b1) $\epsilon_0E_g=-13.6 \times 10^{-3} \,\textrm{C/m}^2$, (a2, b2) $\epsilon_0E_g=0$,
and (a3, b3) $\epsilon_0E_g=13.6 \times 10^{-3}\,\textrm{C/m}^2$.
The bands contributed by the top graphene layer are highlighted by blue color.
The Fermi energy is denoted by the dashed line at $0 \eV $.
Energy shift of molecular HOMO state and the Dirac point ($E_D$) of the top graphene layer by gate voltages is plotted in (c).
}
\end{center}
\end{figure}

\begin{figure}[t]
\begin{center}
\includegraphics[width=\linewidth]{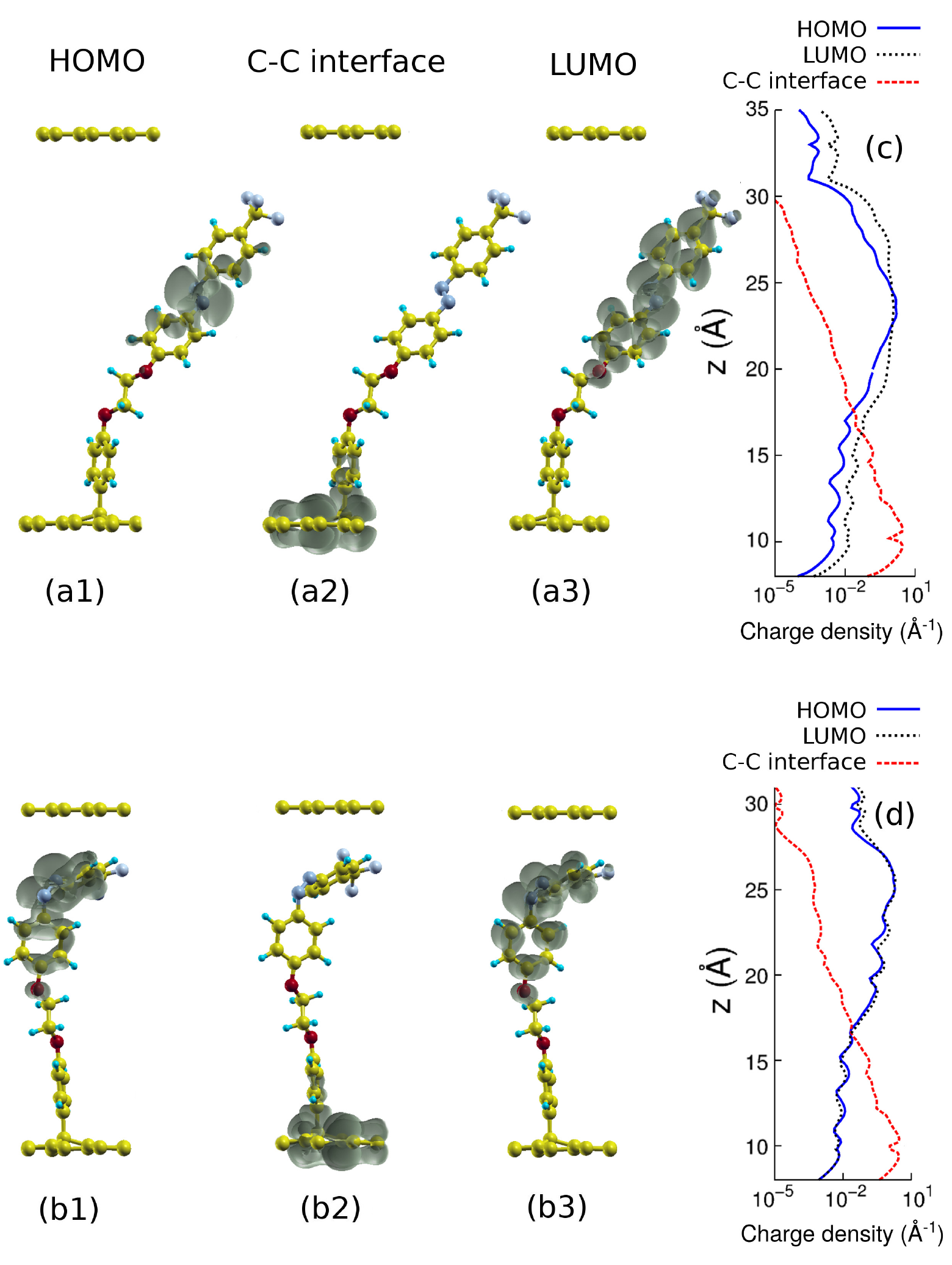}
\caption{
\label{fig:wave-functions}
(color online) 
Charge density corresponding to the (a1,b1) HOMO, (a2,b2) C-C interface state, and (a3,b3) LUMO states
of (a) {\it trans} and (b) {\it cis} junctions.
Charge density integrated over the $x$-$y$ plane (parallel to graphene sheet) along the $z$ direction (normal to graphene sheet)
for (blue solid lines) HOMO, (black dotted lines) LUMO and (red dashed lines) C-C interface states of (c) {\it trans} and (d) {\it cis} junctions.
}
\end{center}
\end{figure}

Do these peaks correspond to states emerging from molecule-graphene or molecule-molecule interactions at the interface? We looked into the electronic structure to identify the states responsible for the transmission peaks near the Fermi energy. With this question in mind we built periodic structures of Gr$\,|\,${\trans}$\,|\,$Gr and Gr$\,|\,${\cis}$\,|\,$Gr whose unit cells are shown in the inset of Fig.~\ref{fig:transport-structure}. Band structures of Gr$\,|\,${\trans}$\,|\,$Gr and Gr$\,|\,${\cis}$\,|\,$Gr are shown in Fig.~\ref{fig:bands}  (a2) and (b2). The strongly dispersive bands spanning the whole energy range come from graphene layers. The bands contributed by the top graphene layer (bonded to molecule via van der Waals interaction)  are highlighted in blue to distinguish them from those of the bottom graphene layer. The Dirac point of both graphene layers is $ 0.15 \eV $ above the Fermi energy, and both of them are hole doped  due to charge transfer with molecules. The molecular orbitals show up as nearly flat bands due to the large distances ($ \sim 7\,\Ang $) among molecules. The LUMO corresponds to the dispersionless band at $1 \eV$  above the Fermi energy, while the HOMO falls around $-0.7 \eV$. The HOMO-LUMO energy gap is $ 1.74 \eV $ for {\trans}  and $1.90 \eV$ for {\cis} molecules. The HOMO and LUMO in both {\trans} and {\cis} molecules are localized on the azobenzene ligand (Fig.~\ref{fig:wave-functions}). For {\trans} molecules, the HOMO is strongly localized between the two benzene rings, while the LUMO spreads over the whole azobenzene ligand [Fig.~\ref{fig:wave-functions}(a1) and (a3)]. The HOMO and LUMO of {\cis} molecules have very similar spatial distributions [Fig.~\ref{fig:wave-functions}(b1) and (b3)].

The state near the Fermi energy can be identified from the band structure. The weakly dispersive bands crossing the Fermi energy [Fig.~\ref{fig:bands} (a2) and (b2)] can be traced to the C-C bond between the molecule and the bottom graphene layer. The charge density corresponding to this state mainly resides on the bottom graphene layer and the phenyl ligand chemically bonded to it; thus it is an interface state [Fig.~\ref{fig:wave-functions}(a2) and (b2)]. The $x$-$y$ plane integrated charge density corresponding to the HOMO, LUMO and the interface state are plotted along the $z$-direction in Fig.~\ref{fig:wave-functions}(c) and (d). The charge density of the interface state decays exponentially across the whole molecule, since its energy is within the LUMO-HOMO gap of the molecule, and the decay rate is faster than both LUMO and HOMO states. The weak dispersion of this state contributes a large density of states at the Fermi energy.


Gate voltages shift the bands of Gr$\,|\,$molecule$\,|\,$Gr junctions. Band structures at gate voltage of $\epsilon_0E_g=-13.6$, $0.0$, and $13.6 \times 10^{-3}\,\textrm{C/m}^2$ are shown in Fig.~\ref{fig:bands}(a,b). The molecular LUMO and HOMO states shift towards lower energy as the gate voltage changes from negative to positive, although the HOMO-LUMO energy gap remains unchanged. The energy shift of bands from the top graphene layer is visible by tracing the position of the Dirac point. These energy shifts are summarized in Fig.~\ref{fig:bands}(c), where the LUMO state is not plotted because of its distance from the HOMO state is unchanged. The effect of gate voltage on the top graphene layer is screened mainly by the bottom graphene layer.
The position of the Dirac point of the top graphene layer in Gr$\,|\,${\trans}$\,|\,$Gr junction shows the same dependence on gate voltage as in Gr$\,|\,${\cis}$\,|\,$Gr junction [Fig.~\ref{fig:bands}(c)].
The energy shift of the HOMO in Gr$\,|\,${\trans}$\,|\,$Gr is weaker than in {\cis} junction, which can be explained by the spatial positions of the molecular orbitals. While the HOMO and LUMO states of {\trans} molecules are near the center of the junction, the molecular states of {\cis} molecules are located away from the bottom graphene layer, so gate voltages have a weaker influence on the orbitals of {\cis} molecules. The band due to the C-C interface state is insensitive to gate voltage because it contributes a large density of states around the Fermi energy. The bands from the bottom graphene layer are pinned by the C-C interface state, so they are also unchanged by gate voltages.

{\it Discussion.} We first discuss the connection between electronic structure and transport properties. Energy shifts of molecular orbitals in response to gate voltages are reflected in the transmission curve. A negative gate voltage raises the HOMO closer to the Fermi energy (Fig.~\ref{fig:bands}). The enhancement of transmission of Gr$\,|\,${\cis}$\,|\,$Gr junctions at negative energies is due to the tail of the Breit-Wigner resonance through the HOMO (Fig.~\ref{fig:transmission_vg}). This phenomenon is not observed in {\trans} junctions because of the strongly localized HOMO. Positive gate voltages lower the  LUMO closer to the Fermi energy, which results in a similar enhancement of transmission at positive energies. This enhancement only happens in the {\cis} junction, in which gate voltages more efficiently tune the energies of molecular orbitals [Fig.~\ref{fig:bands}(c)]. At zero gate voltage, the contribution of the interface state is canceled by the low density of states of the graphene leads. Gate voltage brings the interface state into play by shifting the Dirac point of graphene layer away from the Fermi energy. However, the complexity of the resulting voltage dependence of the transmission curve goes far beyond a simple shift. The interface state contributes peaks in the transmission curve around the Fermi energy and dominates the transmission at positive gate voltage (Fig.~\ref{fig:transmission_vg}).

\begin{figure}[t]
\begin{center}
\includegraphics[width=\linewidth]{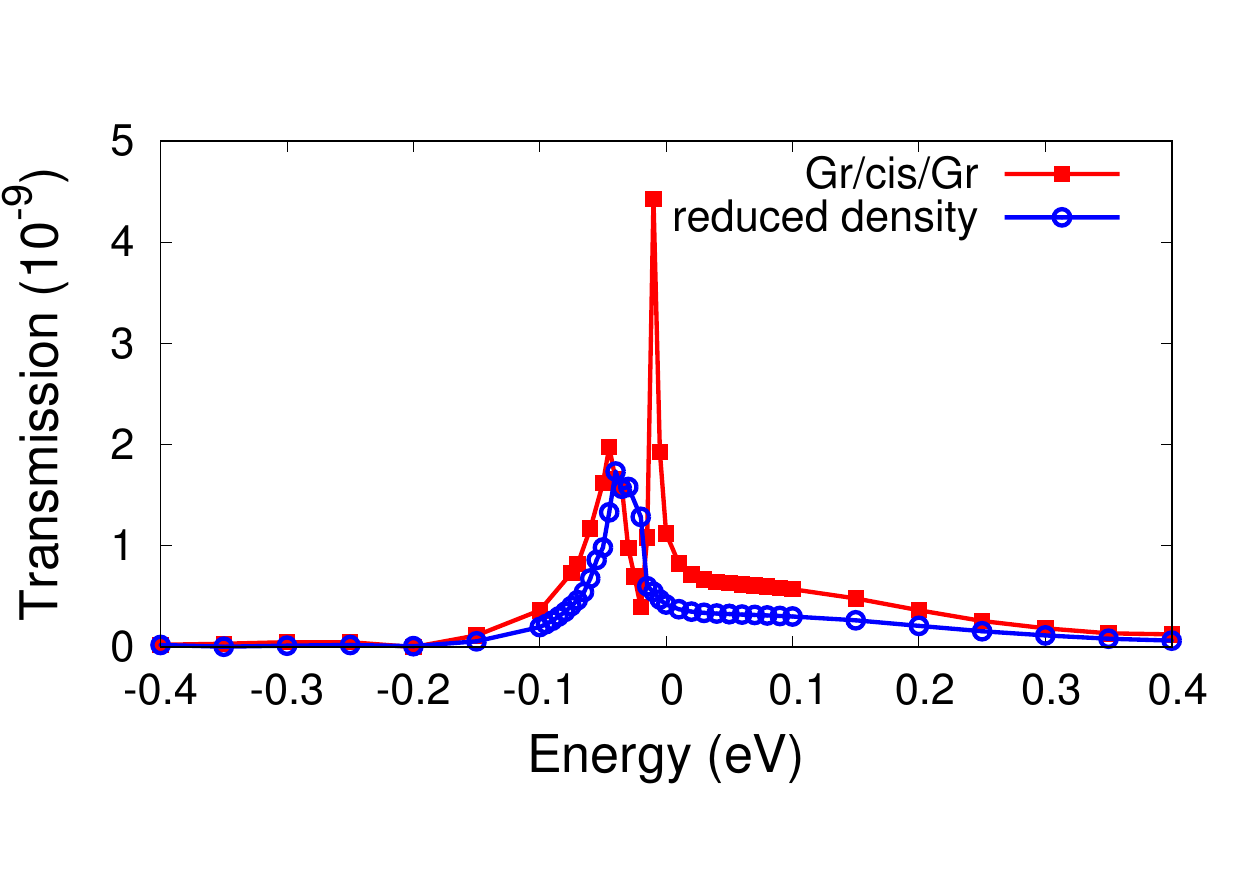}
\caption{
\label{fig:reduced-density}
(color online) 
Transmission of the Gr$\,|\,${\cis}$\,|\,$Gr junction 
and a junction with the density of {\it cis} molecules reduced by half.
The gate voltage is $\epsilon_0 E_g = 12.9 \times 10^{-3} \,\textrm {C/m}^2$.
}
\end{center}
\end{figure}

Next we discuss the mechanism behind the dual-peak in the transmission curve around the Fermi energy at positive gate voltage (top panel, Fig.~\ref{fig:transmission_vg}). At first glance, the asymmetric dual-peak is reminiscent of the Fano resonance \cite{PhysRevB.74.193306,RevModPhys.82.2257,Nat.Nanotech.7.305}, but it was excluded after detailed analysis. The Fano resonance mechanism involves two conducting channels separated in energy: a first (background) channel has an energy far away from the Fano region and a second (local) channel near the Fano region. Interference between the two channels results in an asymmetric transmission peak, and destructive interference can lead to a valley between two peaks. To apply Fano's model to Gr$\,|\,$molecule$\,|\,$Gr junctions, the interface state around the Fermi energy can be seen as the local channel. For the {\trans} junction the molecular LUMO can serve as the background channel, which contributes transmission at positive energies. This explanation however fails for the {\cis} junction, which shows no signature of LUMO contribution to the transmission at positive energies. A novel interpretation is needed. We note that the interface state in Figs.~\ref{fig:wave-functions}(a2)(b2) forms a band with a $ 0.2 \eV $ width [Figs.~\ref{fig:bands}(a,b)] that is of the same order of magnitude as the width of the asymmetric peaks in the transmission. We hypothesize that the asymmetric transmission peak arises from interference of interface states. If true, the width and shape of the transmission peak can be changed by tuning the dispersion and width of the band corresponding to the interface state. We carried out supplementary calculations using the same configuration in Fig.~\ref{fig:transport-structure} but with the density of molecules reduced by half; the band dispersion of the interface state is thus also reduced. Calculations show that the width of the transmission peak is reduced, and more importantly the dual-peak feature disappears (Fig.~\ref{fig:reduced-density}). This supports our hypothesis that the asymmetric transmission peak is contributed by the interface state. This result can also be viewed as an evidence against the Fano mechanism, because none of the global or local channels are killed by reducing the molecular density. 

{\it Conclusion.} 
We studied a vertical field-effect system consisting of graphene and photo-switchable aryl azobenzene molecules. Light irradiation can transform the molecule between {\trans} and {\cis} conformations. Calculations demonstrate the simultaneous control of electronic structure and transport properties by light irradiation and gate voltages. A gate voltage shifts the positions of the LUMO and HOMO of molecules and the bands from graphene. The C-C chemical bonding between molecules and the bottom graphene layer results in an interface state at the Fermi energy. This interface state interactions with other C-C bonding states can lead under finite gate voltage to a strong peak at the Fermi level and therefore dominate the transport properties of junctions. The distinct feature of this junction as shown in the calculated transmission curves (Fig.~\ref{fig:transmission_vg}) illustrates the importance of interfacial processes and predicts that emergent patterns in transmission functions can be engineered.

\begin{acknowledgments}
This work was supported by the US Department of Energy
(DOE), Office of Basic Energy Sciences (BES), under Contract
No. DE-FG02-02ER45995.
Computations were done using the utilities of the
National Energy Research Scientific Computing Center (NERSC).
\end{acknowledgments}



%


\end{document}